\documentclass[%
 reprint,
 amsmath,amssymb,
 mathtools,
 aps,
]{revtex4-1}

\usepackage{graphicx}
\usepackage{dcolumn}
\usepackage{bm}

\def\uu#1{\underline{\underline{#1}}}

\begin{document}

\preprint{APS/123-QED}

\title{The mechanical equilibrium of soft solids with surface elasticity}

\author{Robert W. Style}%
 \email{robert.style@mat.ethz.ch}
\affiliation{Department of Materials, ETH Z\"{u}rich, Switzerland}
\author{Qin Xu}
\affiliation{Department of Materials, ETH Z\"{u}rich, Switzerland}

\date{\today}

\begin{abstract}
Recent experiments have shown that surface stresses in soft materials can have a significant strain-dependence.
Here we explore the implications of this surface elasticity to show how, and when, we expect it to arise.
We develop the appropriate boundary condition, showing that it simplifies significantly in certain cases.
We show that surface elasticity's main role is to effectively stiffen a solid surface's response to in-plane tractions, in particular at length-scales smaller than a characteristic elastocapillary length.
We also investigate how surface elasticity effects the Green's-function problem of a line force on a flat, linear-elastic substrate.
There are significant changes to this solution, especially in that the well-known displacement singularity is regularised.
This raises interesting implications for soft phenomena like wetting contact lines, adhesion and friction.
Finally, we discuss open questions, future directions, and close ties with existing fields of research.

\end{abstract}

\pacs{Valid PACS appear here}
\maketitle


\section{Introduction}

Solid surface stresses are forces that emerge at surfaces and interfaces in solids, in analogy to the concept of surface tension in liquids.
Typically they are overlooked, as they are too weak to cause any observable deformations.
However recently, a large body of experimental evidence has emerged showing that, in softer solids like polymer gels, surface stresses can significantly change material behaviour and properties \cite{mora11,jago12,chak13,nade13,park14,karp15,mond15,andr16,styl17,bico18}.
Examples include how surface stresses modify wetting on soft surfaces \cite{styl13,styl13b,nade13,karp15}, change adhesive behaviour of small particles on soft substrates \cite{styl13c,jens15,ina17}, and can even play a leading role in determining the stiffness of soft composites \cite{ducl14,styl15}.
This gives a basis to theoretical work that has long predicted how surface stresses might influence physical phenomena (e.g. \cite{shut50,gurt78,camm94,mill00,spae00,shar04,ding05,duan05,shen05,bris10,manc17}).

However, recent experiments suggest that surface stresses may be more complex than is typically assumed \cite{xu17,jens17,schu17}:
so far, almost every experimental work has treated these as taking a uniform, isotropic value, just like the surface tension of a liquid (e.g. \cite{kund09,jago12,chak13,mora13,nade13,styl13,karp15,mond15}).
This is a natural assumption, as soft solids -- where elastocapillary effects are mostly observed -- are often polymer gels.
These typically have a large solvent component, and so their surfaces might be expected to behave like that of the pure solvent (e.g. \cite{hui13}).
However, recent results have shown that surface stresses can be very strain-dependent, and that this strain-dependence can significantly affect soft-material behaviour.
First, experiments measured the shape of a wetting ridge under a contact line on a soft silicone surface, and used this to show that surface stresses approximately double when the gel is stretched biaxially by 20\% \cite{xu17}.
Second, adhesion experiments measured the forces during the pull-off of a silica bead (with radius $\sim 10\mu$m) from a soft silicone gel \cite{jens17}.
These forces could be split into three parts: one due to each of the bulk elastic response, the isotropic surface tension of the unstretched solid, and the strain-dependant surface elasticity.
Each of these were of similar magnitude, and thus equally important to the adhesion process.

Motivated by these experiments, here we explore the implications of surface elasticity.
The equations governing surface stresses depend heavily on the local surface geometry.
However in several situations they simplify, giving us intuition into the role of surface elasticity, and insight into how and when we expect it to arise.
We show how surface-elastic phenomena depend heavily on characteristic elastocapillary lengthscales.
We also show how surface elasticity regularises the singularity for the classic problem of a line force acting on the surface of a linear-elastic solid, and discuss the implications for the equilibrium of a pinned contact line on a soft surface.
Finally, we discuss important, new directions for understanding surface-influenced phenomena in soft materials.
 
\section{The surface stress boundary condition}

Surface stresses are surface forces that arise parallel to material surfaces or interfaces.
These occur due to changes in the molecular structure of materials in the close vicinity of surfaces from their bulk molecular structure.
For example, at simple liquid surfaces, the molecular density decreases within a few nanometres of the surface.
The molecules there try to pull back together, and the resulting tensile forces appear as an isotropic, strain-independent surface tension at the macroscopic level \cite{jone99}.

\begin{figure}
\centering
\includegraphics[width=1\linewidth]{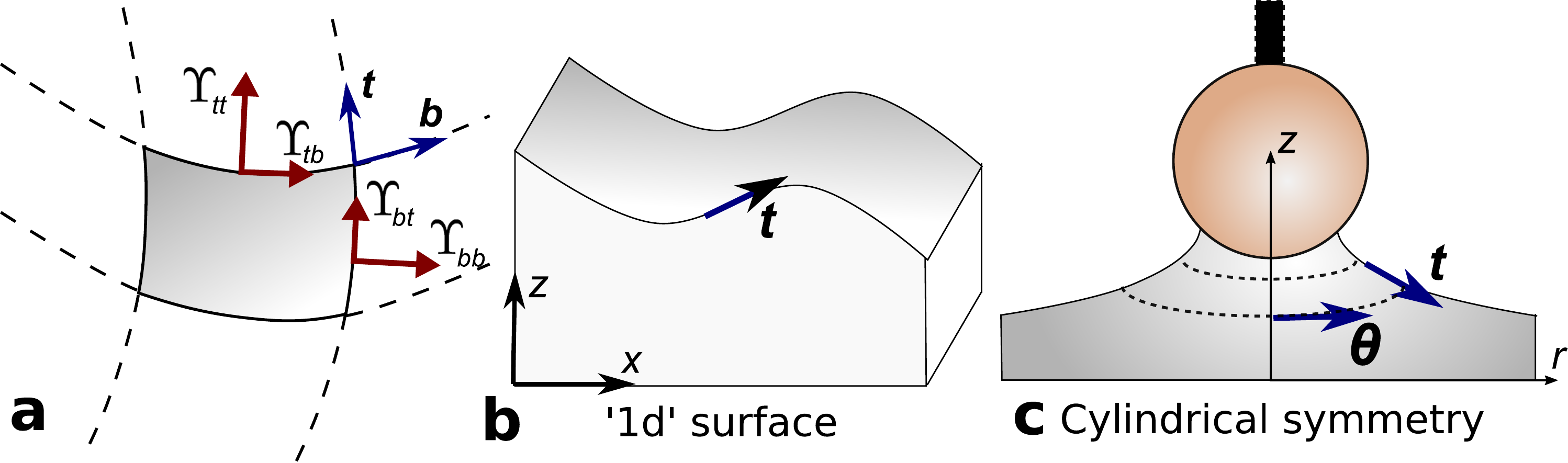}
\caption{a) The components of surface stress acting on a small section of a material's surface. $\underline{t}$ and $\underline{b}$ are perpendicular, tangent vectors to the surface. b,c) Schematics showing the geometry for a  translationally invariant surface, and a surface with cylindrical symmetry respectively.}
\label{fig:schematic}
\end{figure}

Similar changes in the molecular structure at the surface of solids causes surface stresses to appear.
These are not necessarily isotropic, but instead can be represented by a local, 2x2 surface-stress tensor, $\uu{\Upsilon}$ -- effectively a surface version of the Cauchy stress tensor, $\uu{\sigma}$ (e.g. Figure \ref{fig:schematic}a).
A useful image of this is to imagine a very thin, stiff, elastic sheet with no bending rigidity, that is stretched out and attached to the surface of a solid.
Then, at the macroscopic scale, the stresses in the sheet would appear as apparent surface stresses.

The mechanical effect of surface stresses is to cause a jump in the stresses across a solid interface.
This is encapsulated in the surface-stress boundary condition,
\begin{equation}
[\underline{\underline{\sigma}}.\underline{n}]^+_-=-\underline{\nabla}^s.\underline{\underline{\Upsilon}},
\label{eqn:bc}
\end{equation}
that relates the jump in the normal component of the bulk stresses, $\uu{\sigma}$, to the surface stresses.
Here, $\underline{n}$ is the normal unit vector to the surface, $\underline{\nabla}^s$ is the surface gradient operator \cite{styl17}, and $[x]^+_-$ indicates the jump in $x$ across the interface.

For liquid-like materials with an isotropic surface stress, 
$\uu{\Upsilon}=\Upsilon \uu{I}$ (where \uu{I} is the 2D, surface identity tensor), this becomes:
\begin{equation}
[\underline{\underline{\sigma}}.\underline{n}]^+_-=\Upsilon {\cal K}\underline{n}-\underline{\nabla^s \Upsilon},
\label{eqn:liq_like}
\end{equation}
where ${\cal K}=\underline{\nabla}^s.\underline{n}$ is the total curvature.
This is a familiar expression from liquid capillarity:
the first, normal term on the right-hand side is the Laplace pressure jump across a curved interface.
the second, tangential term is the equivalent of a Marangoni stress.

When $\uu{\Upsilon}$ is anisotropic, it still makes sense to decompose equation (\ref{eqn:bc}) into normal and tangential parts.
The normal component can always be written as a generalised Laplace pressure jump:
\begin{equation}
-\underline{n}.(\underline{\nabla}^s.\underline{\underline{\Upsilon}})=\underline{\underline{\Upsilon}}:\underline{\underline{\nabla^s n}}\equiv\underline{\underline{\Upsilon}}:\underline{\underline{\cal{K}}},
\end{equation}
where $\underline{\underline{\cal{K}}}$ is the curvature tensor of the surface, and $\Upsilon:{\cal K}=\Upsilon_{ij}{\cal K}_{ij}$.
However, the form of the tangential component of equation (\ref{eqn:liq_like}) is not so easy to interpret physically. If $\underline{t}$ is a tangent vector, then
\begin{equation}
-\underline{t}.(\underline{\nabla}^s.\underline{\underline{\Upsilon}})=-\underline{\nabla^s}.(\underline{t}.\uu{\Upsilon})+\underline{\underline{\Upsilon}}:\underline{\underline{\nabla^s t}}.
\end{equation}
Thus, there is always a term that depends on gradients in the surface stress (like Marangoni stresses), and a term that depends on the local coordinate system chosen. This is typically very complex for general surfaces \cite{duan09}. However, it does simplify in some useful cases.

First, in the case of a `1d' surface like that in Figure \ref{fig:schematic}b, where the material is translationally invariant along one direction (e.g. for the case of a long, straight contact line on a soft solid), we can write $\uu{\Upsilon}=\uu{\Upsilon}(t)$, where $t$ is the arclength co-ordinate.
Then we find 
\begin{equation}
[\underline{\underline{\sigma}}.\underline{n}]^+_-=\Upsilon_{tt}{\cal K}\underline{n} - \frac{d\Upsilon_{tt}}{d t}\underline{t},
\label{eqn:2dsurf}
\end{equation}
where $\Upsilon_{tt}(t)=\underline{t}.\uu{\Upsilon}.\underline{t}$.
Thus we recover a boundary condition that is completely equivalent to the liquid-like case in equation (\ref{eqn:liq_like}).

Second, when there is cylindrical symmetry around a central axis (such as for a sphere probe adhering to a soft surface, as shown in Figure \ref{fig:schematic}c), the surface can again be parameterised in terms of arclength, $t$, so that $\uu{\Upsilon}=\uu{\Upsilon}(t)$.
In this case
\begin{equation}
[\underline{\underline{\sigma}}.\underline{n}]^+_-=\Upsilon_{tt}{\cal K}\underline{n} - \left(\frac{d\Upsilon_{tt}}{d t} + (\Upsilon_{tt}-\Upsilon_{\theta\theta})\underline{\nabla}^s.\underline{t} \right)\underline{t},
\end{equation}
where now $\Upsilon_{\theta\theta}$ is the tensile surface stress in the azimuthal direction.
Thus we seem a similar form emerge to the previous case, but with an extra term that depends on the divergence of the coordinate system.
Earlier, we mentioned the analogy between surface stress, and the idea of a thin, elastic sheet at the interface between two phases.
This is further borne out by noting that, for the case of an infinitely soft solid, the equation above becomes that used to model a pendant droplet coated in a thin elastic sheet \cite{knoc13,nage17}, or the indentation of a floating elastic sheet \cite{box17}.

The final step to being able to address mechanical problems involving surface stress is to choose a constitutive  relation between $\uu{\Upsilon}$ and the surface strain, $\uu{\epsilon}^s$.
The natural choice is the linear-elastic constitutive relationship:
\begin{equation}
\uu{\Upsilon}=\Upsilon_0 \uu{I}+2\mu^s \uu{\epsilon}^s+\lambda^s \mathrm{Tr}(\uu{\epsilon}^s) \uu{I}.
\label{eqn:surf_const_reln}
\end{equation}
In fact, recent experiments have shown that, for soft silicone gels, this holds very well up to strains of more than 30\% \cite{xu17,xu18}:

\section{Tractions applied to a flat substrate with surface elasticity}

We now focus on a translationally invariant system like that in Figure \ref{fig:schematic}c in which we can make useful analytic progress.
We consider an initially-flat, linear-elastic film of thickness $h$ on a rigid substrate, apply tractions, $\underline{\tau}(x)$ to the surface, and calculate the resulting surface displacements.
The film has Young's modulus, $E$, and Poisson's ratio $\nu$, and the local displacements of the film, $\underline{u}=(u_x,u_z)$ satisfy the static equilibrium:
\begin{equation}
(1-2\nu)\nabla^2 \underline{u}+\underline{\nabla}(\underline{\nabla}.\underline{u})=0,
\end{equation}
as well as the linear-elastic constitutive relation.

For the boundary condition, we linearise equations (\ref{eqn:2dsurf},\ref{eqn:surf_const_reln}) to find that
\begin{equation}
\underline{\tau}-\underline{\underline{\sigma}}.\underline{z}=-\frac{\partial^2}{\partial x^2}\left( \begin{matrix}
\Lambda  u_x \\
\Upsilon_0 u_z
 \end{matrix}\right)
\label{eqn:lin_bc}
\end{equation}
Here $\Lambda=\lambda^s+2\mu^s$, and $\underline{z}$ is the unit vector in the $z$ direction.
Note that we immediately see that the undeformed surface tension, $\Upsilon_0$, affects the normal stress balance at the surface, while surface elasticity, $\Lambda$, affects the tangential stress balance.

This problem can be solved by moving into Fourier space in the x-coordinate (e.g. \cite{jeri11,xu10,styl12}). Then,
\begin{equation}
\sigma_{iz}(k,h)=Q_{ij}u_j(k,h),
\end{equation}
where $\uu{Q}(k,h)$ is given in \cite{xu10,jeri11}.
Additionally, equation (\ref{eqn:lin_bc}) becomes
\begin{equation}
\tau_i(k)-\sigma_{iz}(k,h)=S_{ij}u_j(k,h),
\end{equation}
where $\uu{S}=((\Lambda k^2 \,\, 0), (0\,\, \Upsilon_0k^2))$. Thus
\begin{equation}
u_i(k,h)=QS^{-1}_{ij}\tau_j(k)
\label{eqn:QS}
\end{equation}
where $\uu{QS}=\uu{Q}+\uu{S}$, and we can calculate the surface displacements for any given traction distribution.

For an incompressible solid with $\nu=1/2$ (which is generally assumed to be a good approximation for gels and elastomers),
\begin{widetext}
\begin{equation}
QS_{xx}^{-1}=\frac{h}{E}\frac{1}{\bar{k}^2(\bar{\Upsilon}_0+\bar{\Lambda})-\frac{\bar{k}(-4-8\bar{k}^2 +24 \bar{k}^2\bar{\Upsilon}_0-9\bar{k}^2\bar{\Upsilon}_0^2(1+2 \bar{k}^2)-(4-9\bar{k}^2\bar{\Upsilon}_0^2)\cosh{2\bar{k}}   ) }{3(4\bar{k}-3\bar{k}\bar{\Upsilon}_0-6 \bar{k}^3\bar{\Upsilon}_0 + 3 \bar{k}\bar{\Upsilon}_0\cosh{2\bar{k}}+2\sinh{2\bar{k}} )}},
\label{eqn:QSxx}
\end{equation}
\begin{equation}
QS_{xz}^{-1}=-QS_{zx}^{-1}=\frac{h}{E}\frac{12 i \bar{k}}{\left(4+8\bar{k}^2-9\bar{k}^2(1+2\bar{k}^2)\bar{\Upsilon}_0\bar{\Lambda}+12 \bar{k}^2(\bar{\Lambda}-\bar{\Upsilon}_0)+(4+9\bar{k}^2 \bar{\Lambda}\bar{\Upsilon}_0)\cosh{2\bar{k}}+ 6\bar{k}(\bar{\Lambda}+\bar{\Upsilon}_0)\sinh{2\bar{k}} \right)}
\end{equation}
\begin{equation}
QS_{zz}^{-1}=-\frac{h}{E\bar{k}}\frac{3\bar{k}(4+3\bar{\Lambda}+6\bar{k}^2\bar{\Lambda}-3\bar{\Lambda}\cosh{2\bar{k}})-6\sinh{2\bar{k}} }{\left(4+8\bar{k}^2-9\bar{k}^2(1+2\bar{k}^2)\bar{\Upsilon}_0\bar{\Lambda}+12 \bar{k}^2(\bar{\Lambda}-\bar{\Upsilon}_0)+(4+9\bar{k}^2 \bar{\Lambda}\bar{\Upsilon}_0)\cosh{2\bar{k}}+ 6\bar{k}(\bar{\Lambda}+\bar{\Upsilon}_0)\sinh{2\bar{k}} \right)}.
\label{eqn:QSzz}
\end{equation}
\end{widetext}
Here, $i=\sqrt{-1}$, and we have used non-dimensional parameters: $\bar{k}=kh$, $\bar{\Upsilon}_0=\Upsilon_0/Eh$ and $\bar{\Lambda}=\Lambda/Eh$.

\section{The line-force problem}

\begin{figure}[t]
\centering
\includegraphics[width=0.9\linewidth]{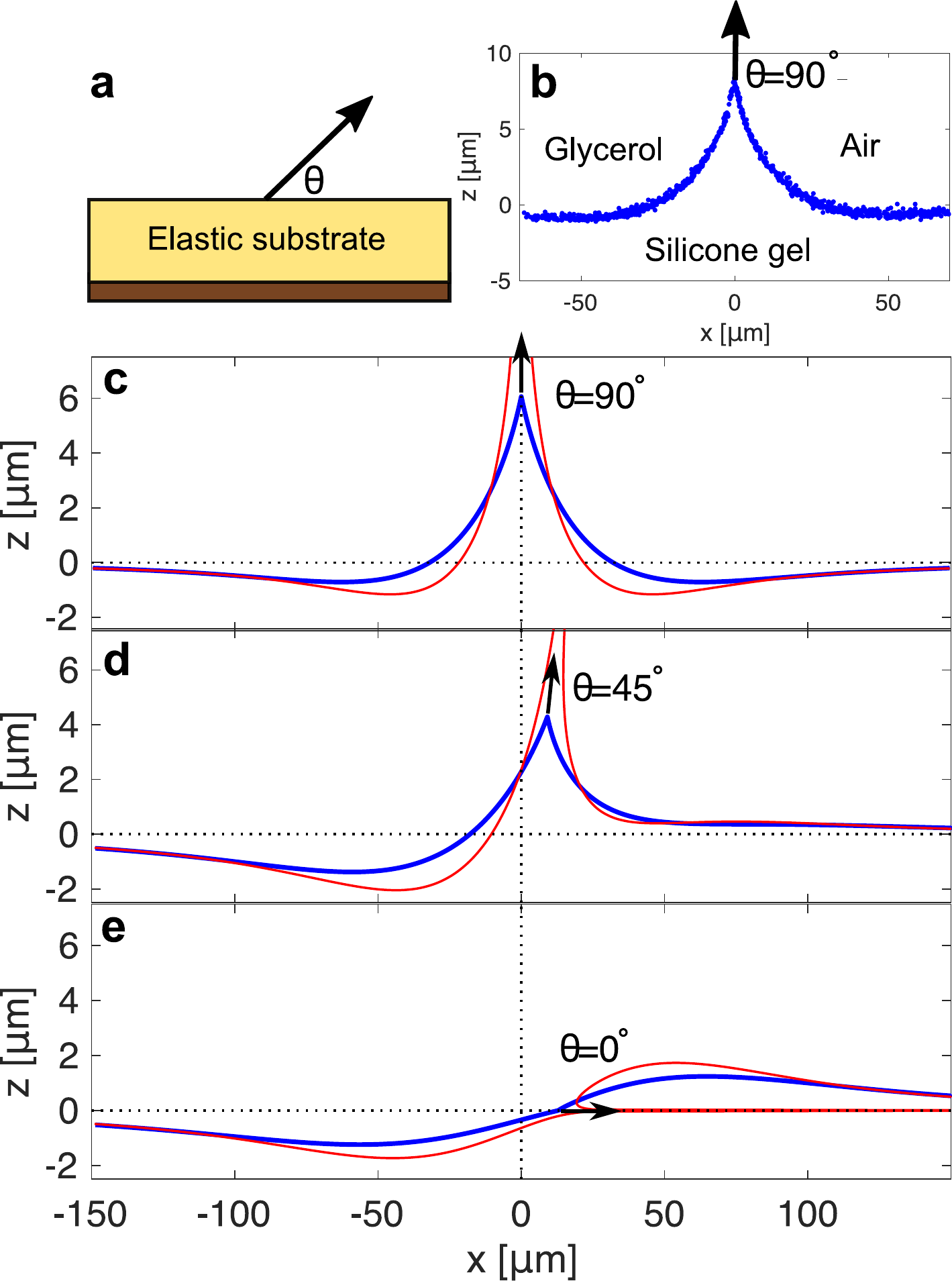}
\caption{The effect of surface stresses on the shape of wetting ridges at contact lines. a) Schematic diagram. b) Example experimental data for the surface profile of a $50\mu$m-thick, silicone-gel substrate under the contact line of a glycerol droplet with $\theta=90^\circ$ (taken from the data set of \cite{styl13}. Here $E=3$kPa. c-e). Calculated surface profiles for contact lines with $\theta=90^\circ,45^\circ$ and $0^\circ$ respectively. Thick, blue curves and thin, red curves correspond to calculations with/without surface stresses respectively. Here, we take $E=3$kPa, $\nu=1/2$, $\gamma_l=\Upsilon_0=\Lambda=0.03$N/m and $h=50\mu$m. For the case of no surface stresses, $\Upsilon_0=\Lambda=0$, and we clearly see the displacement singularities at the contact line.}
\label{fig:surface_profiles}
\end{figure}

We use this solution to study the problem of a line force at an angle on a soft, flat substrate (Figure \ref{fig:surface_profiles}a).
This is important for two reasons.
First, this is equivalent to the problem of a straight, pinned, droplet contact line on a soft, flat substrate.
In this case, the surface is pulled up to form a wetting ridge, like that shown in Figure \ref{fig:surface_profiles}b.
Second, the resulting surface profiles form the basis of the Green's function approach for the problem of tractions applied to such a substrate.
In other words, we can build up solutions to a general traction problem by adding up the surface responses to many line forces distributed at different positions along the surface.

Mathematically, this problem is equivalent to setting $\underline{\tau}=\gamma_l \delta(x) (\cos \theta, \sin\theta)$, where $\delta(x)$ is the delta function, $\theta$ is the droplet's contact angle, and $\gamma_{l}$ is the droplet's surface tension.
We assume that the surface has the same constitutive equation (\ref{eqn:surf_const_reln}) either side of the contact line.

In the case of no surface stresses, this problem is simply the classical line-force problem of linear elasticity, which is well-known to have a displacement singularity at the contact line (e.g. \cite{shan86,jeri11}).
Previously, we have shown that when $\uu{\Upsilon}$ is a constant, liquid-like surface tension, the singularity vanishes, but only for the case that $\theta=90^\circ$ \cite{jeri11,styl12}.
Here, we show that the addition of surface elasticity also eliminates the singularity for all $\theta$.

To solve the problem, we note that in Fourier space, $\tau(k)=\gamma_l (\cos\theta,\sin\theta)$. Thus equation (\ref{eqn:QS}) gives:
\begin{equation}
\underline{u}(x)=\frac{\gamma_l}{2\pi} \int_{-\infty}^{\infty} e^{ikx}  \uu{QS}^{-1}.\begin{bmatrix}
    \cos\theta \\
    \sin\theta 
\end{bmatrix}  dk.
\end{equation}
The results are given by the thick, blue curves in Figure \ref{fig:surface_profiles}(c-e) for some typical parameters for soft gels ($\gamma_l=\Upsilon_0=\Lambda=0.03$N/m, $h=50\mu\mathrm{m}$, $\nu=1/2$ and $E=3\mathrm{kPa}$), and for three different contact angles: $\theta=0,45,90^\circ$.
Additionally we plot the surface displacements for the case of no surface stresses: $\Upsilon_0=\Lambda=0$, as shown by the thin, red curves.

We can immediately make several observations.
First, surface stress eliminates the displacement singularity for all $\theta$ -- not just $90^\circ$.
We can see why this occurs by integrating equation (\ref{eqn:lin_bc}) with respect to $x$ from $0^-$ to $0^+$.
In the in-plane direction, the delta function in $\underline{\tau}$ is now balanced by a jump in $\Lambda \partial u_x/\partial x$ at the contact line, rather than by the elastic stresses in the substrate.
Similarly, in the out-of-plane direction, the delta function is balanced by a jump in $\Upsilon_0 \partial u_z/\partial x$ -- observable as a sharp corner at the wetting ridge tip.

Second, there is a significant difference between the wetting-ridge shapes with and without surface stress.
The difference appears across almost the whole width ridge, and is not just confined to the ridge tip.
Thus there would be a significant error if were to use the surface-stress-free solutions as a basis for a Green's function approach.
In this case we would lose a lot of accuracy, especially at small scales.
For example, for the surface stress parameters in Figure \ref{fig:surface_profiles}(c-e), we would not be able to rely on predicted features with a horizontal size $\lesssim100\mu$m.

Third, this solution allows a detailed investigation of how a droplet can achieve mechanical equilibrium at the contact line, even if the contact line is pinned.
There are several interesting questions that immediately arise.
For example, horizontal force balance relies on the presence of surface elasticity.
Thus, do we expect to see very different macroscopic pinning, hysteresis, or dynamic contact line \cite{extr96,carr96,karp15} behaviour between two materials with different surface elastic constants?
If a material has no surface elasticity, how does pinning occur, and what is the role of nonlinear elasticity and plasticity at the ridge tip?

\begin{figure}
\centering
\includegraphics[width=1\linewidth]{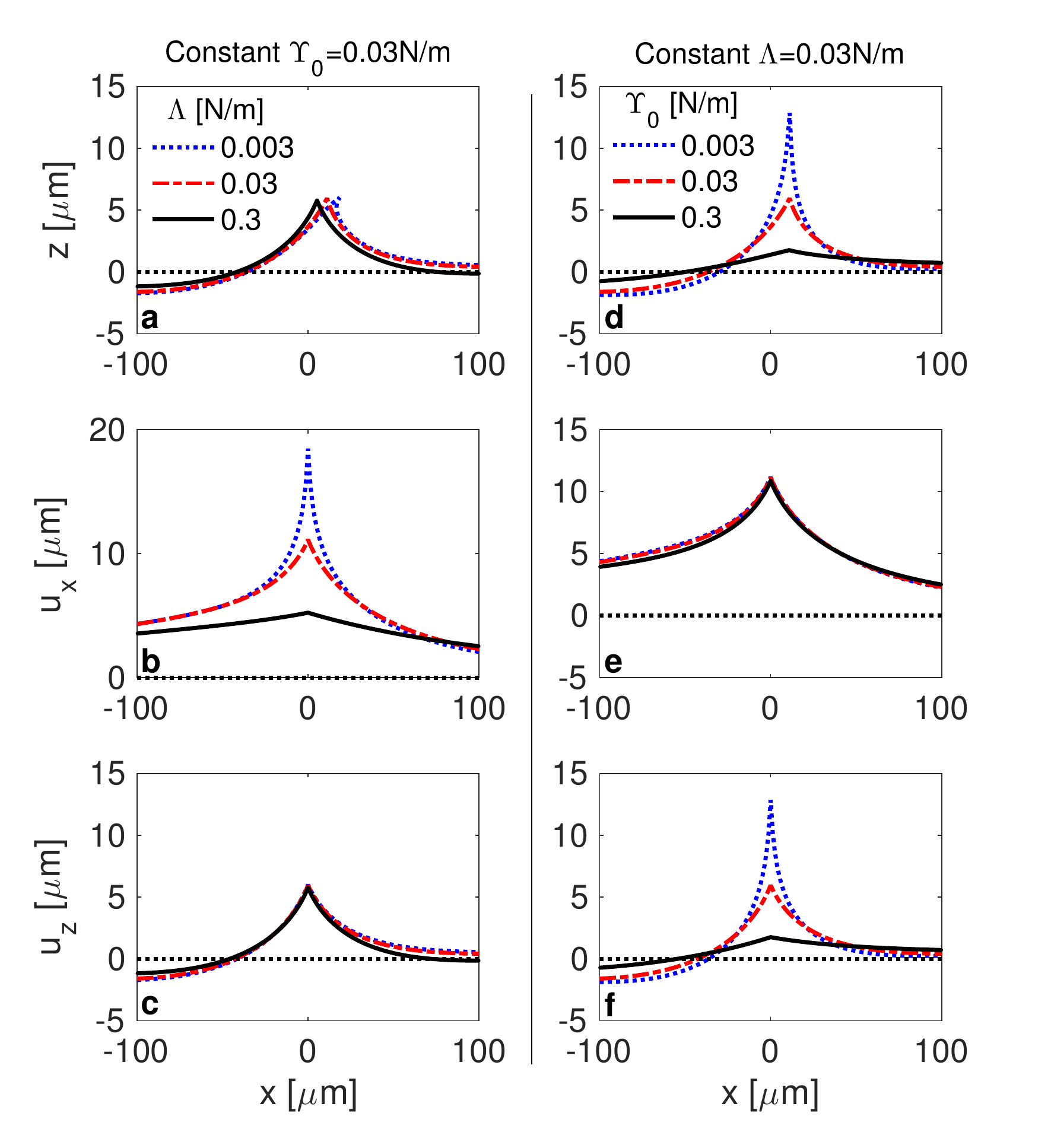}
\caption{The effect of changing surface elasticity, $\Lambda$, and the undeformed surface tension, $\Upsilon_0$, on the solution for a line force applied at $45^\circ$ to the surface of a solid substrate. The top row shows the resulting surface profiles, while in the second and third rows, we break this down into the $x$ and $z$ surface displacements respectively.
We set $E=3$kPa, $\nu=1/2$, $h=50\mu$m, $\gamma_l=0.03$N/m and $\theta=45^\circ$.
In a-c), we hold $\Upsilon_0$ constant, and vary $\Lambda$. This strongly affects $u_x$ (b), but makes very little difference to $u_z$ (c).
In d-f) we hold $\Lambda$ constant and vary $\Upsilon_0$. In this case, only $u_z$ changes significantly (f). This illustrates how $\Lambda$ and $\Upsilon_0$ predominantly affect in-/out-of-plane surface response respectively.}
\label{fig:different_roles}
\end{figure}

Earlier, we noted that $\Lambda$ affects shear stresses at the surface, while $\Upsilon_0$ affects normal stresses (from equation \ref{eqn:lin_bc}).
Similarly, we find that for a given applied traction field, $\Lambda$ and $\Upsilon_0$ largely affect only the resulting in-plane/out-of-plane surface displacements respectively.
This is illustrated in Figure \ref{fig:different_roles}, which shows both in-plane (second row) and out-of-plane (third row) surface displacements for a contact line on a substrate with $\theta=45^\circ$.
All parameters except $\Upsilon_0$ and $\Lambda$ are kept the same as before.
In the left column of the Figure, we vary $\Lambda$ by two orders of magnitude while keeping $\Upsilon_0$ constant.
In this case, the $u_x$ profile is a strong function of $\Lambda$, while the $u_z$ profile stays almost constant.
Similarly, in the right column, we vary $\Upsilon_0$ while keeping $\Lambda$ constant.
Here, $u_x$ barely changes, while $u_z$ is a strong function of $\Upsilon_0$.
This further suggests that the main role of $\Lambda$ is to impart in-plane stiffness to the surface, while $\Upsilon_0$ mainly opposes out-of-plane forces.

We also note that the results of Figure \ref{fig:different_roles} agree with the behaviour seen in recent experimental results by Xu et al. \cite{xu18}.
They measured the shape of a wetting ridge, as the substrate was stretched perpendicular to the contact line.
In this case, the angle of the ridge tip broadened with increasing stretch.
Within our model, applying a substrate pre-stretch with strain, $\epsilon_\infty$, effectively changes $\Upsilon_0\rightarrow \Upsilon_0 + \Lambda \epsilon_\infty$.
Thus we can mimic the experiments by exploring how changing $\Upsilon_0$ changes the surface profile.
Indeed Figure \ref{fig:different_roles}(d) shows that increasing $\Upsilon_0$ causes the angle of the ridge tip to increase.

\section{When do surface stress effects arise?}

The analysis above suggests that surface elasticity can strongly affect a surface's  response to traction forces, but when do they become significant?
This information is held within the matrix $\uu{QS}^{-1}(k,h,\Upsilon_0,\Lambda)$, which is effectively the surface compliance for a given wavenumber $k$.
In particular, the largest (and thus most important) contributions come from $QS^{-1}_{xx}$ and $QS^{-1}_{zz}$, which represent the in-plane surface response to in-plane tractions, and the out-of-plane surface response to out-of-plane tractions respectively.

\begin{figure}
\centering
\includegraphics[width=1.1\linewidth]{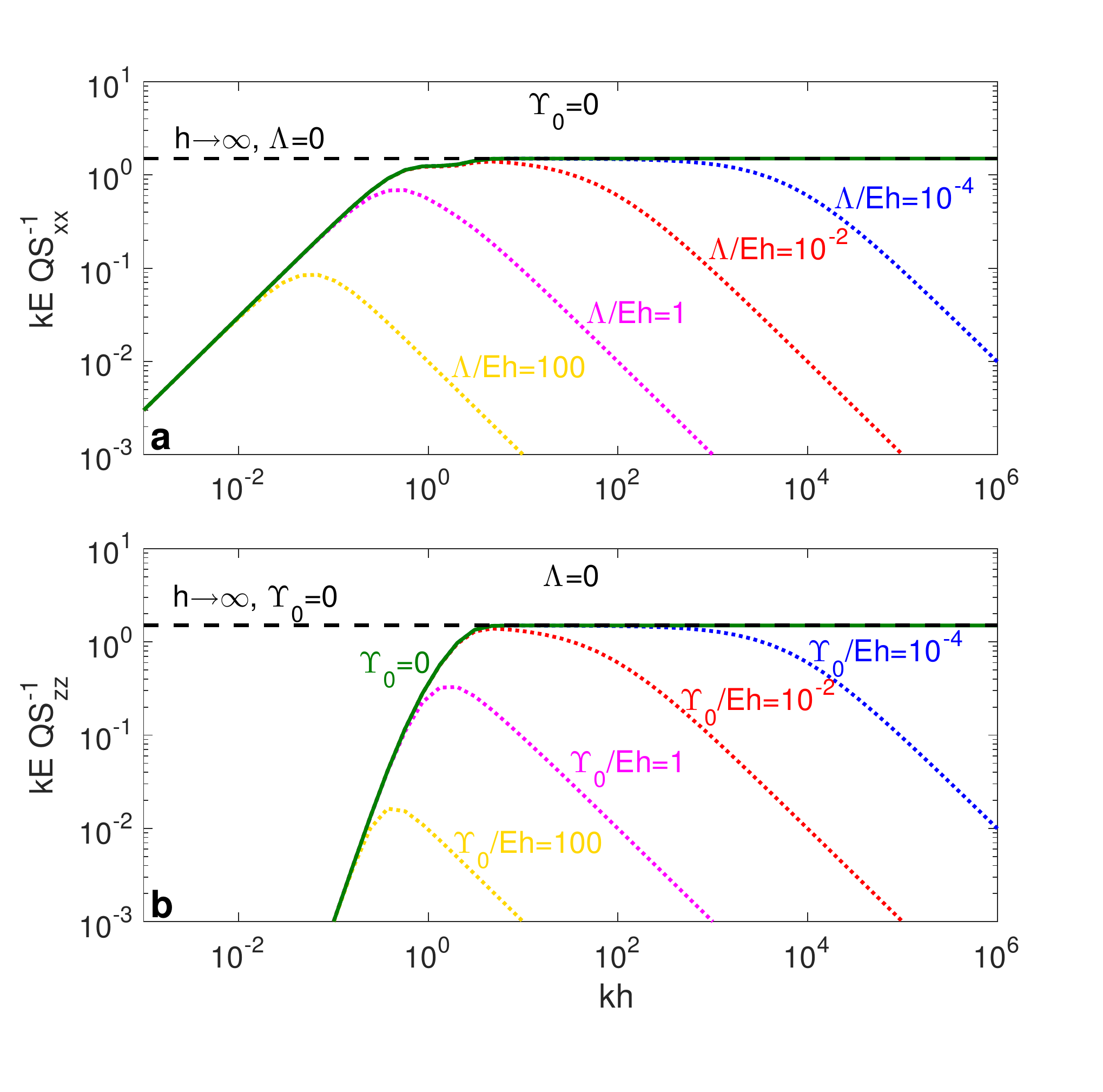}
\caption{The (non-dimensionalised) in-plane (a) and out-of-plane (b) surface compliances as a function of (non-dimensional) wavenumber $kh$. Black, dashed lines show that the surface compliance is wavelength-independent for an infinitely deep substrate with no surface stress. Green, continuous curves show how this is altered by the presence of a rigid bottom boundary. This reduces surface compliance for long wavelengths (small $kh$). The dashed curves show the effects of changing surface elasticity, $\Lambda$ (a), and the undeformed surface tension, $\Upsilon_0$ (b). These reduce surface compliance for small wavelength (large $kh$).}
\label{fig:damping}
\end{figure}

Figure \ref{fig:damping}(a,b) shows how $QS^{-1}_{xx}$ and $QS^{-1}_{zz}$ depend on $\bar{k}=kh$ for various different parameter values.
Note that in the plots, we non-dimensionalise $QS^{-1}_{xx}$ and $QS^{-1}_{zz}$ by multiplying them by $kE$.
Figure \ref{fig:damping}(a) shows the in-plane compliance, $QS^{-1}_{xx}$, for a variety of different values of the non-dimensional surface elasticity, $\Lambda/Eh$.
The form of these curves changes very little with $\Upsilon_0$ (see Appendix), so we choose to set it to be zero.
Similarly Figure \ref{fig:damping}(b) shows the out-of-plane compliance, $QS^{-1}_{zz}$, for a variety of different values of the non-dimensional, undeformed surface tension $\Upsilon_0/Eh$.
These curves change very little with $\Lambda$ (see Appendix), so we choose it to be zero here.

The plots in Figure \ref{fig:damping} highlight when surface displacements are reduced by confinement or surface stresses.
In both plots, the dashed black lines show the surface compliance for an infinitely thick substrate with no surface stress.
This is a constant value (there is no natural lengthscale in this case that would allow the surface response to be a function of $k$).
The green curves show the surface compliance for a confined elastic layer with thickness $h$, and no surface stresses.
In this case, the response is the same as for an infinitely thick substrate for $kh\gtrsim 1$, but the surface compliance is lower at longer wavelengths when $1/k\equiv\lambda\lesssim h$, due to the presence of the rigid bottom of the substrate.
The dotted curves show the surface compliance in the presence of surface stresses, and how this is reduced for large $k$, short wavelength perturbations.

The equations (\ref{eqn:QSxx},\ref{eqn:QSzz}) are rather complicated, but can be simply approximated to gain insight into the physics.
In the short-wavelength limit $\bar{k}\gg 1$, we find that
\begin{equation}
QS^{-1}_{xx}=\left(\frac{2Ek}{3}+\Lambda k^2\right)^{-1},\,\, QS^{-1}_{zz}=\left(\frac{2Ek}{3}+\Upsilon_0 k^2\right)^{-1}
\end{equation}
In the other, long-wavelength limit, $\bar{k}\ll 1$,
\begin{equation}
QS^{-1}_{xx}=3h/E,\,\, QS^{-1}_{zz}=h^3 k^2/E.
\end{equation}
Note that, for a contact line, the first, surface-stress-dominated, limit dictates the small-scale structure of a wetting ridge close to a contact line.
The second, confinement-dominated, limit dictates the large-scale response of a surface away from the wetting ridge.
We combine the two limits, in the same way one calculates the total resistance of resistors in parallel, to make the approximations:
\begin{equation}
QS^{-1}_{xx}\approx \frac{1}{\frac{E}{3h}+\frac{2Ek}{3}+\Lambda k^2},
\label{QSxx_simp}
\end{equation}
\begin{equation}
QS^{-1}_{zz}\approx \frac{1}{\frac{E}{h^3k^2}+\frac{2Ek}{3}+\Upsilon_0 k^2}.
\label{QSzz_simp}
\end{equation}
These show very good agreement with the full numerical solution, as shown in the Appendix.

The expressions (\ref{QSxx_simp},\ref{QSzz_simp}) clearly demonstrate the essential physics of the problem.
First, they again show how $\Lambda$ and $\Upsilon_0$ affect predominantly in-plane and out-of-plane substrate compliances respectively.
Second, the factors in their denominators represent the effect of the three factors that oppose surface deformations: confinement from the rigid bottom, bulk elasticity, and surface stress respectively.
By examining the magnitude of each of these terms, we see when surface stress becomes important.
For in-plane stresses, there are two separate cases, depending on the relative sizes of the elastocapillary length $\Lambda/E$ and $h$.
In the case of a deep substrate with $h\gtrsim \Lambda/E$, surface-stress effects arise for wavelengths $\lesssim \Lambda/E$.
On the other hand, for a shallow substrate with $h\lesssim \Lambda/E$, surface-stress effects arise for wavelengths $\lesssim \sqrt{\Lambda h/E}$.

For out-of-plane stresses, we draw similar conclusions, depending on the relative sizes of $\Upsilon_0/E$ and $h$.
For deep substrates with $h\gtrsim \Upsilon_0/E$, surface-stress effects arise for wavelengths $\lesssim \Upsilon_0/E$.
For shallow substrates with $h\lesssim \Upsilon_0/E$, surface-stress effects arise for wavelengths $\lesssim (\Upsilon_0 h^3/E)^{1/4}$.

\section{Discussion \& Conclusions}

The results above show how surface stresses are important at small lengthscales in soft materials. 
It is well established that a constant, isotropic surface tension plays an important role in soft materials at small lengthscales (e.g. \cite{styl17}).
However, here we have also shown how surface elasticity is expected to dominate the small-scale response of soft materials to in-plane tractions.
For an idea of typical values, measurements in soft silicone gels give $\Lambda=83.3$mN/m and $E=3$kPa \cite{xu18}, so in this case we expect surface elasticity to become important at scales $\lesssim \Lambda/E=30\mu$m.
Our results suggest that many processes involving tractions may need to be reanalysed to account for surface elasticity.
Such processes include friction involving small patches of contact and contact-line pinning and dynamics.

Although the model presented above provides insight into the essential physics of surface stresses, it is a simplified geometry, and there remains much work to understand the role of surface stresses in more general systems.
However, we note that it may be possible to draw from established techniques in other, related fields.
For example, for a 2d linear-elastic solid, the equations of motion are very similar to those for viscous flow (e.g. \cite{pali90}), while the governing stress-boundary condition (\ref{eqn:2dsurf}) is identical.
Thus it may be possible to derive solutions to the equations using techniques used for studying Marangoni flows in viscous liquids (e.g. \cite{land88}).
Similarly, the surface layer behaves, mathematically, exactly like a thin elastic sheet with no bending rigidity.
Thus, it may be possible to adapt techniques from membrane/vesicle science, and from the study of thin-elastic sheets, or wrinkling of bilayers (e.g. \cite{dank09,stoo15,box17}).

One key application where further work is needed is adhesion.
Experiments have shown that the adhesive forces between a small indenter and a soft, silicone-gel surface depended strongly on surface elasticity \cite{jens17}.
An important challenge is thus to be able to make quantitative predictions in order to work with similar, soft systems.
We should note that, at first sight, the experimental observations appear to contradict the results above, i.e. that surface elasticity does not affect out-of-plane forces.
However, this is likely because substrate strains were very large in the experiments, and thus the linearising assumptions used above clearly do not hold.
To capture such observations, potential approaches could include techniques like:
(i) not linearising the surface stress boundary condition (\ref{eqn:lin_bc}), (ii) adapting approaches that have been used from elastic sheet theory to predict indentation forces in the large deformation limit (e.g. \cite{box17}), or (iii) matching techniques, where the problem is divided into a surface-stress-dominated inner region near the indenter, that is matched onto an outer region that behaves like the linear model above.

It is worth noting that in the discussion above, we have made the continuum approximation that all scales in the problem are much larger than a characteristic molecular dimension ($\sim$1\AA).
Following previous work, we expect that the physics described above will be modified when the elastocapillary lengths, $\Upsilon_0/E$ and $\Lambda/E$ approach such a scale \cite{marc12b}, and what occurs in this case is also an interesting question for further exploration.

In conclusion, we have explored the role of surface elasticity in soft solids.
At leading-order, this role is to reduce the compliance of a surface to in-plane tractions, in particular at length-scales below a characteristic elastocapillary length.
This means that many processes involving small-scale traction forces, friction and adhesion on soft surfaces may behave significantly differently to what would be expected from existing theory.
Surface elasticity also regularises the singularity in the displacement field for the problem of a straight, pinned contact line on a soft, linear-elastic substrate, raising interesting questions about both static and dynamic wetting problems on soft surfaces.
Moving forward, we hope we have demonstrated that there are many close connections with existing fields of research, and interesting open questions that make this a rich area for both theoretical and experimental research.

\section{Appendix}

\begin{figure}
\centering
\includegraphics[width=1.1\linewidth]{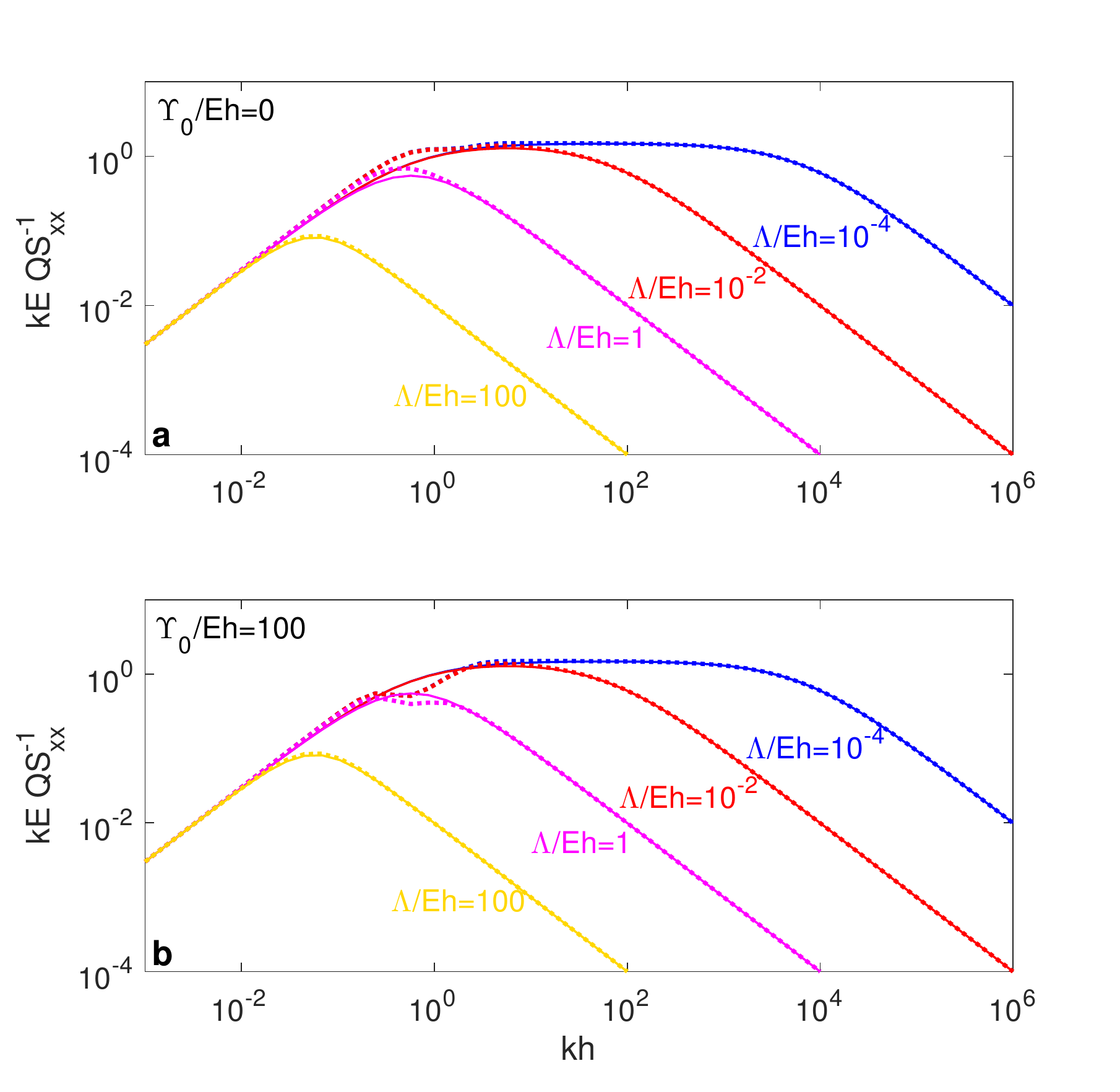}
\caption{The (non-dimensionalised) in-plane compliance (dotted curves) is relatively independent of the undeformed surface tension, $\Upsilon_0$. This is shown by plotting $QS_{xx}^{-1}$ (Equation (\ref{eqn:QSxx})) for various different values of $\Lambda/Eh$, for the two cases $\Upsilon_0/Eh=0$ (a, dotted curves) and $\Upsilon_0/Eh=100$ (b, dotted curves). There is very little difference between the two, illustrating that the in-plane compliance has a far stronger dependence on the value of $\Lambda/Eh$ than $\Upsilon_0/Eh$. We also show the approximation (\ref{QSxx_simp}) for each set of parameters (continuous curves), finding good agreement with the full expressions.}
\label{fig:QSxx_appendix}
\end{figure}

\begin{figure}
\centering
\includegraphics[width=1.1\linewidth]{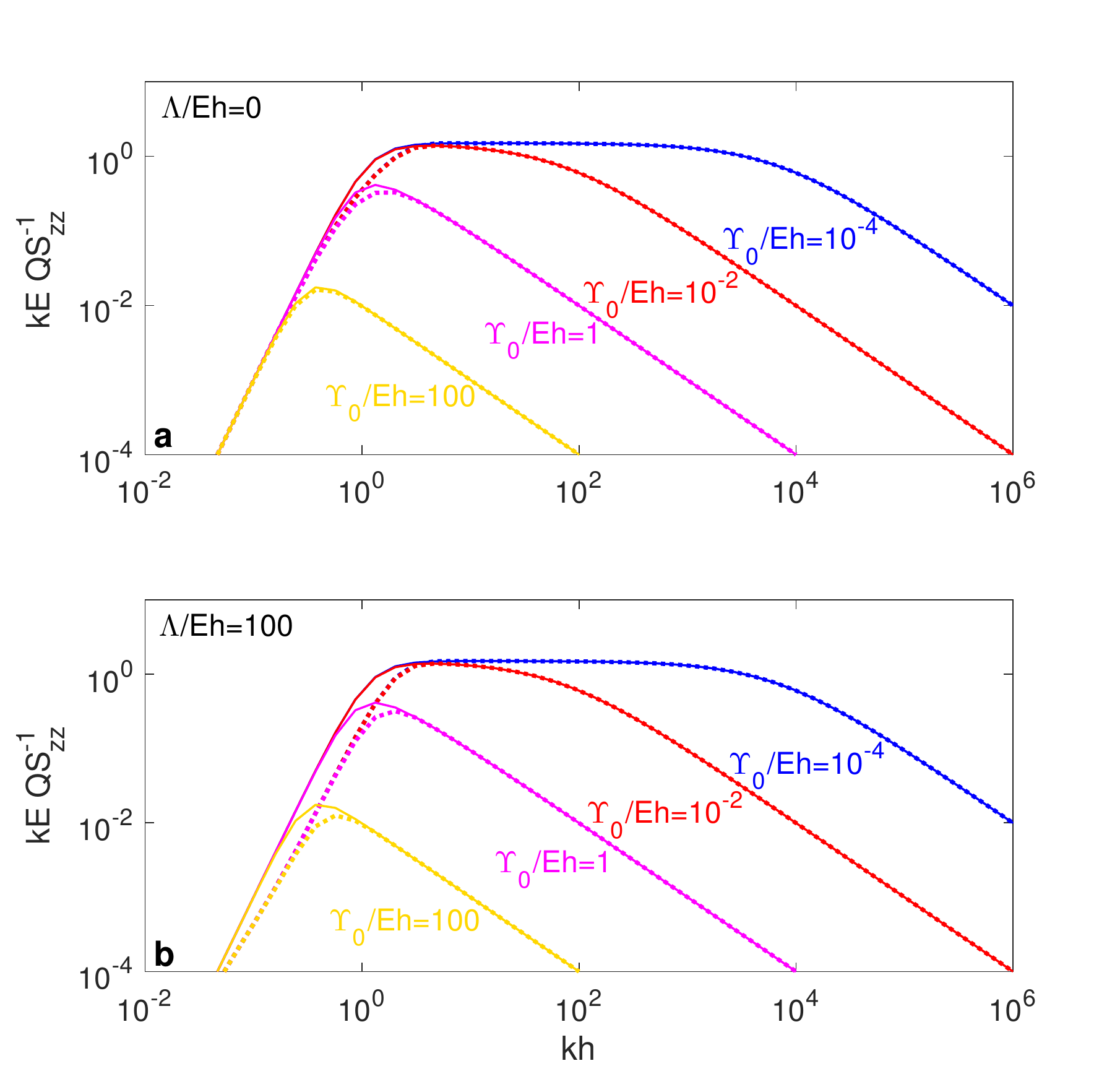}
\caption{The (non-dimensionalised) out-of-plane compliance (dotted curves) is relatively independent of the undeformed surface tension, $\Lambda$. This is shown by plotting $QS_{zz}^{-1}$ (Equation (\ref{eqn:QSzz})) for various different values of $\Upsilon/Eh$, for the two cases $\Lambda/Eh=0$ (a, dotted curves) and $\Lambda/Eh=100$ (b, dotted curves). There is very little difference between the two, illustrating that the out-of-plane compliance has a far stronger dependence on the value of $\Upsilon_0/Eh$ than $\Lambda/Eh$. We also show the approximation (\ref{QSzz_simp}) for each set of parameters (continuous curves), finding good agreement with the full expressions.}
\label{fig:QSzz_appendix}
\end{figure}

Here, we give a few further details about the compliances $QS^{-1}_{xx}$ and $QS^{-1}_{zz}$.
Firstly we show that these are relatively independent of $\Upsilon_0$ and $\Lambda$ respectively, as illustrated in Figures \ref{fig:QSxx_appendix},\ref{fig:QSzz_appendix}.
In Figure \ref{fig:QSxx_appendix}, $QS_{xx}^{-1}$ is shown by the dotted curves for a range of values of $\Lambda/Eh$ in two cases: firstly when $\Upsilon_0/Eh=0$ (top) and secondly when $\Upsilon_0/Eh=100$ (bottom).
Despite the very large increase in $\Upsilon_0/Eh$, there is very little difference between these two sets of curves -- $QS_{xx}^{-1}$ is much more sensitive to changes in $\Lambda/Eh$.
Similarly, in Figure \ref{fig:QSzz_appendix}, $QS_{zz}^{-1}$ is shown by the dotted curves for a range of values of $\Upsilon_0/Eh$.
There is very little difference between the cases $\Lambda/Eh=0$ (top) and $\Lambda/Eh=100$ (bottom).

The figures also demonstrate the accuracy of the approximations (\ref{QSxx_simp},\ref{QSzz_simp}).
In both figures, these are shown as the continuous curves.
Aside from some small deviations for $kh\sim 1$, the approximations show excellent agreement with the full expressions for $QS^{-1}_{xx}$ and $QS^{-1}_{zz}$.

\section*{Conflicts of interest}
There are no conflicts to declare.

\section*{Acknowledgements}
We thank Eric Dufresne for useful discussions. RWS is partially funded by SNSF Grant 200021-172827.

\end{document}